\documentclass[letter,11pt,twoside]{article}

\usepackage[dvipdfmx]{graphicx,color}
\usepackage[dvipsnames]{xcolor}
\usepackage{graphbox,subcaption}
\usepackage{amsmath,amssymb,mathrsfs,bm,bold-extra}
\usepackage[TU]{fontenc}
\usepackage{newtxtext, newtxmath}
\usepackage{tabularray,booktabs,multirow,dcolumn,bigdelim}
\usepackage{arydshln}
\usepackage{fancybox, mdframed}
\usepackage{enumitem,sectsty,titlesec}
\usepackage[hang,flushmargin]{footmisc}
\usepackage{endnotes}
\usepackage{pdflscape,afterpage,capt-of}
\usepackage{fancyhdr}
\usepackage{setspace}
\usepackage{here,stackengine,empheq,cases,braket}
\usepackage[makeroom]{cancel}
\usepackage{mwe,ulem}
\usepackage{filecontents,etoolbox}
\usepackage{comment}
\usepackage[compress,authoryear,round]{natbib}
\usepackage[font={footnotesize,stretch=1.1},labelsep=space]{caption}
\usepackage{subcaption}
\DeclareCaptionLabelFormat{vert}{\textbf{{#1} {#2}~|}}
\usepackage[hyphens]{xurl}
\usepackage[pdfencoding=auto,linkbordercolor=CornflowerBlue,citecolor=magenta,linkcolor=tb,pdfpagelabels=false]{hyperref}
\hypersetup{colorlinks=true,allbordercolors=CornflowerBlue}

\usepackage{orcidlink}

\renewcommand*{\thepage}{\footnotesize\arabic{page}}
\renewcommand*{\thefootnote}{\textcolor{black}{$\myfnsymbol{\value{footnote}}$}}
\makeatletter
\newcommand*{\myfnsymbol}[1]{\ensuremath{%
\ifcase#1 \or \ast \or \dagger \or \ddagger \or \spadesuit \or \diamondsuit \or \clubsuit \or \heartsuit \else \@ctrerr \fi}}
\makeatother

\definecolor{tb}{rgb}{0.24, 0.43, 0.91}

\def\mtitle{Evolving interdisciplinary contributions to global societal challenges: A 50-year overview}

\begin{document}

\quad\vspace{-2.0cm}
\begin{flushright}
October 2025~~[v2]
\end{flushright}

\vspace{0.5cm}
\begin{center}
\fontsize{13pt}{14pt}\selectfont\bfseries

       \scalebox{0.96}{\mtitle}            %
\end{center}

\vspace*{0.8cm}

\centerline{%
{Keisuke Okamura}\,\footnote{\,\tt 
okamura@alumni.lse.ac.uk}%
\orcidlink{0000-0002-0988-6392}\mbox{\small ${}^{\,1,\,2}$}}

\vspace*{0.6cm}
{\small
\centerline{\textit{%
${}^{1}$Ministry of Education, Culture, Sports, Science and Technology (MEXT),}}
\centerline{\textit{%
3-2-2 Kasumigaseki, Chiyoda-ku, Tokyo 100-8959, Japan.}}
\vspace*{3mm}
\centerline{\textit{%
${}^{2}$SciREX Center, National Graduate Institute for Policy Studies (GRIPS),}}
\centerline{\textit{%
7-22-1 Roppongi, Minato-ku, Tokyo 106-8677, Japan.}}
\vspace*{0.5cm}}

\vspace{1.0cm}
\noindent\textbf{Abstract.}
\quad
Addressing global societal challenges necessitates insights and expertise that transcend the boundaries of individual disciplines.
In recent decades, interdisciplinary collaboration has been recognised as a vital driver of innovation and effective problem-solving, with the potential to profoundly influence policy and practice worldwide.
However, quantitative evidence remains limited regarding how cross-disciplinary efforts contribute to societal challenges, as well as the evolving roles and relevance of specific disciplines in addressing these issues.
To fill this gap, this study examines the long-term evolution of interdisciplinary contributions to the United Nations' Sustainable Development Goals (SDGs), drawing on extensive bibliometric data from OpenAlex.
By analysing publication and citation trends across 19 research fields from 1970 to 2022, we reveal how the relative presence of different disciplines in addressing particular SDGs has shifted over time.
Our results also provide unique evidence of the increasing interconnection between fields since the 2000s, coinciding with the United Nations' initiative to tackle global societal challenges through interdisciplinary efforts.
These insights will benefit policymakers and practitioners as they reflect on past progress and plan for future action, particularly with the SDG target deadline approaching in the next five years.

\vspace{0.8cm}
\vfill

\thispagestyle{empty}
\setcounter{page}{1}
\setcounter{footnote}{0}
\setcounter{figure}{0}
\setcounter{table}{0}
\setcounter{equation}{0}

\newpage
\renewcommand{\thefootnote}{\arabic{footnote}}
\setlength{\skip\footins}{10mm}
\setlength{\footnotesep}{4mm}
\let\oldheadrule\headrule
\renewcommand{\headrule}{\color{VioletRed}\oldheadrule}

\pagestyle{fancy}
\fancyhead[LE,RO]{\textcolor{VioletRed}{\footnotesize{\textsf{\leftmark}}}}
\fancyhead[RE,LO]{}
\fancyfoot[RE,LO]{\color[rgb]{0.04, 0.73, 0.71}{}}
\fancyfoot[LE,RO]{\scriptsize{\textbf{\textsf{\thepage}}}}
\fancyfoot[C]{}
\thispagestyle{empty}

\newpage
\tableofcontents
\vfill

\begin{mdframed}[linecolor=magenta]
\begin{tabular}{l}
\begin{minipage}{0.14\hsize}
\hspace{-1.5mm}\includegraphics[width=1.8cm,clip]{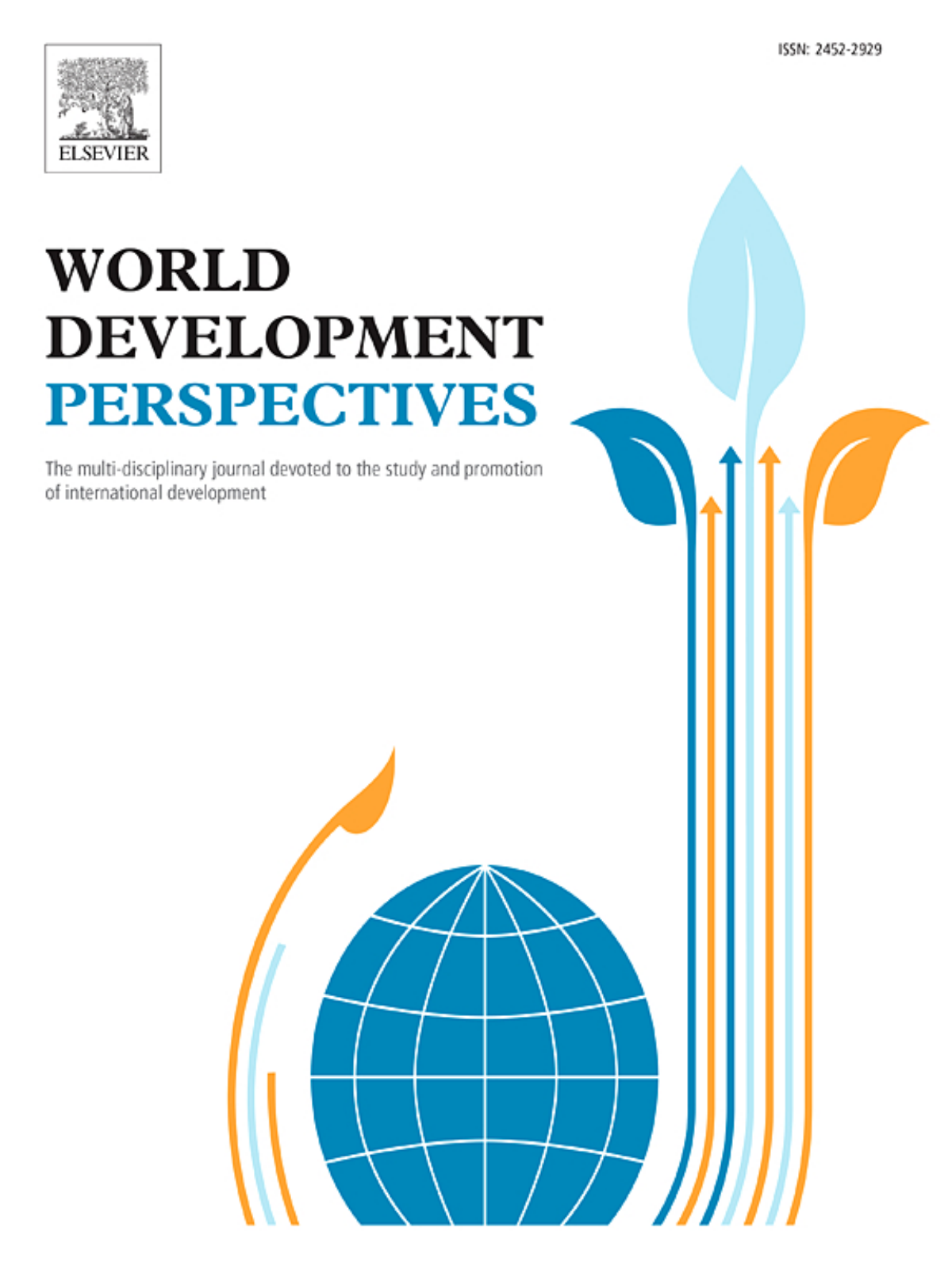}
\end{minipage}
\begin{minipage}{0.83\hsize}
\setstretch{0.95}
\textcolor{magenta}{\footnotesize\textrm{This version reproduces the content of the paper authored by K.~Okamura, titled \textit{`{\mtitle}'}, published in \textit{World Development Perspectives} (doi: \href{https://doi.org/10.1016/j.wdp.2025.100728}{10.1016/j.wdp.2025.100728}).
The supplementary materials accompanying this manuscript have been seamlessly integrated into this single file.}}
\end{minipage}
\end{tabular}
\end{mdframed}

\newpage
\section{Introduction: Interdisciplinary efforts for global societal challenges\label{sec:intro}}

It is widely recognised that the insights and expertise of a single academic discipline are insufficient to address global challenges effectively.
Consequently, the importance of integrating knowledge and efforts across disciplines is increasingly acknowledged as essential to tackling these challenges.
A substantial body of prior research has demonstrated the pivotal role of interdisciplinary collaboration in fostering innovation and addressing societal problems \citep{Adler18,Hu24,Ledford15,Repko20}.
Over the past few decades, numerous policy documents from national and international governmental organisations \citep{EU15,Gleed16}, alongside initiatives from academia \citep{Aldrich14,NAS05}, and industry \citep{Kotiranta20,Giri23}, have drawn on these experiences and lessons to shape their objectives and strategies.
Notably, at the Millennium Summit in September 2000, the United Nations (UN) Millennium Declaration led to the establishment of eight Millennium Development Goals (MDGs) \citep{UN00}.
This was followed in September 2015 by the adoption of 17 Sustainable Development Goals (SDGs) at the UN Summit on Sustainable Development \citep{UN15}.
Both frameworks are grounded in cross-disciplinary principles, motivations, and approaches.

\begin{figure*}[!t]
\centering
\vspace{-0.5em}
\includegraphics[width=0.74\textwidth]{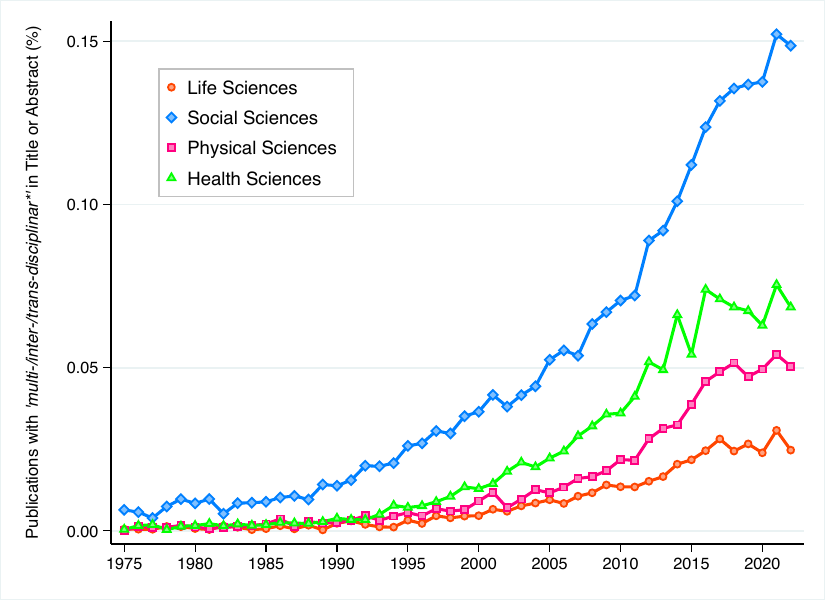}
\caption{\textbf{The number of publications related to `interdisciplinarity' by publication year and domain.} [Source: OpenAlex]}\label{fig:1}
\end{figure*}

In response to these global trends, the visibility and impact of publications related to interdisciplinarity have expanded markedly in recent years.
Figure \ref{fig:1} illustrates the proportion of publications indexed in OpenAlex \citep{Priem22}---one of the largest catalogues of the world's scholarly research---that contain the terms `\textit{multi(-)disciplinary}', `\textit{inter(-)disciplinary}', or `\textit{trans(-)disciplinary}' in their titles or abstracts, plotted by year of publication.
The plots were generated for each of the four broad field categories, referred to as domains in OpenAlex.
The graph highlights a noteworthy increase in the prevalence of these terms within the Social Sciences over the past few decades.
The other three domains---Physical Sciences, Health Sciences, and Life Sciences---show a generally consistent upward trend, albeit with some variation in absolute numbers.
This finding aligns with similar analyses conducted a decade ago \citep{Lariviere14,VanNoorden15}.

Nevertheless, there remains a lack of systematic and long-term quantitative evidence on how interdisciplinary efforts have contributed to addressing global societal challenges.
Prior bibliometric studies have mapped linkages between societal challenges and scientific subfields, profiled journal-level participation, and compared national contributions to the SDGs \citep{ElsSDG23,Garai23,Mishra24,Raman24,Sweileh20,Yamaguchi23,Yumnam24}.
Collectively, this literature delineates where disciplinary engagement is concentrated versus diffuse and how SDG-relevant outputs are distributed across fields and venues.
However, most analyses focus on specific domains or limited time windows, providing only partial evidence on long-run, cross-SDG interdisciplinarity.

The knowledge and strategies effective in addressing specific societal challenges can differ markedly depending on the nature of the issue.
It is entirely possible for individuals---researchers and policymakers alike---from completely different disciplines and policy areas to work on the same challenge without recognising each other's contributions, applying knowledge and methods unique to their respective fields.
For instance, in the case of SDG 5, which aims to achieve gender equality and empower all women and girls, disciplines such as sociology, political science, or psychology may immediately come to mind.
However, as shown later (Figure \ref{fig:3}), medicine has also made a substantial contribution in recent years, a fact not readily apparent when compared with the aforementioned fields.

Each societal challenge therefore often involves significant contributions from disciplines that might not initially be expected.
Recognising these contributions and their temporal dynamics provides a vital perspective on how interdisciplinary efforts evolve and how they contribute to addressing global challenges.
Such an understanding also helps reveal the organic collaborative relationships and structures between academia and the policy sphere that work towards common global problems.
These insights are invaluable for shaping future collaboration across disciplines, sectors and borders.

In practice, conducting such analyses has proven challenging, primarily due to the difficulty of assessing whether, and to what extent, a given contribution has helped address, let alone resolve, a societal challenge \citep{DEste23, Sandes22}.
These evaluations are inherently part of a complex social and political process, and the methods, timeframes, and social consensus required make it hard to discuss the results with any degree of certainty---scientifically, politically or practically.
Nevertheless, in today's era of rapidly advancing Open Science and Data Science \citep{Dong17,Wittenburg21}, and with the increasing availability of diverse scientometric data, it has become more feasible than ever to gain indirect insights through extensive data analysis.
This is the approach we adopt in this paper.

While previous studies provide valuable insights into the relationships among scientific subfields, journals, and national or international contributions to the SDGs, important gaps remain.
Most notably, we still lack systematic, longitudinal evidence on how disciplinary contributions to global societal challenges have evolved over time and the extent to which different challenges follow distinct trajectories of interdisciplinarity across the pre-MDG and post-SDG periods.
Against this backdrop, the present study addresses the following guiding research question:
How have the connections between research disciplines and global societal challenges, as represented by the SDGs, developed over time?

To answer this question, the paper proceeds as follows.
Section \ref{sec:method} describes the data and methodology, focusing on two machine learning–based measures derived from the OpenAlex platform: the field-affinity scores and the SDG-affinity scores.
Section \ref{sec:result} presents the main results, highlighting both overall trends and SDG-specific dynamics.
Section \ref{sec:concl} discusses the implications of these findings and concludes with a synthesis of the study’s contributions, limitations, and directions for future research.

\section{Methods: Tracing interdisciplinary contributions to SDGs via bibliometrics\label{sec:method}}

\subsection{The OpenAlex data}

The data used in the current study were retrieved from OpenAlex \citep{Priem22} between January and April 2024.
OpenAlex gathers information on various types of publications, including journal articles, non-journal articles, reviews, preprints, conference papers, books, and datasets.
This large-scale data catalogue employs the latest artificial intelligence (AI) technologies to analyse the disciplines closely related to various societal challenges and how these relationships have evolved over time.
As a focal point for global societal challenges, we adopt the UN's SDGs, introduced earlier, as a prominent and contemporary framework.
Using bibliometric methods, we explore how different disciplines in the natural and social sciences have collectively addressed or related to these SDGs.

For disciplines, we use the 19 level-0 fields from the `concept' taxonomy in OpenAlex.
These fields include: Political Science, Philosophy, Economics, Business, Psychology, Mathematics, Medicine, Biology, Computer Science, Geology, Chemistry, Art, Sociology, Engineering, Geography, History, Materials Science, Physics, and Environmental Science, collectively referred to as $\mathscr{D}$ hereafter.\footnote{From 2024 onwards, OpenAlex has introduced new classification categories to replace `concepts', including 4,516 `topics' (\url{https://help.openalex.org/how-it-works/topics}), 252 `subfields', 26 `fields', and 4 `domains'.
For the analysis in the current paper, however, we chose to continue using `concepts' because it offers an optimal level of granularity for our purposes---neither too detailed with too many fields nor too broad with too few fields---allowing for meaningful comparisons across a wide range of disciplines.}
For each of these fields, we collected the metadata for the top 1,000 most-cited publications from each publication year between 1970 and 2022.
It is important to note that selecting publications based on citation counts does not imply that these publications are inherently superior in terms of scientific merit.
Rather, we use citation counts as a proxy indicator of attention, reflecting the level of interest these publications have garnered from researchers and the wider community.

In total, the dataset comprised $N=100{,}700$ individual publications.
This number results from sampling 1,000 publications per field per publication year, across 19 fields ($\mathscr{D}$) and 53 publication years (1970--2022).
Refer to Supplementary Materials, including Tables \ref{tab:1} and \ref{tab:2}, for further details on the OpenAlex data used in this study.

\subsection{The field-affinity score}

The metadata for each publication in OpenAlex includes a score indicating how closely the publication relates to the 19 level-0 concepts,\footnote{\url{https://docs.openalex.org/api-entities/works/work-object\#concepts}.} which we refer to as the field-affinity score.
These scores are calculated using specific deep learning algorithms \citep{Barrett23}.
The field-affinity score is assigned to each publication for each field as a number between 0 and 1, with 0 meaning the publication is unrelated to the field and 1 indicating maximal relevance to the field.
Using the field-affinity scores, we calculated the distances between the 19 level-0 concepts ($\mathscr{D}$) for each publication year ($y$), following the same procedure outlined in \cite{Okamura19}, as follows.

To calculate the distance between fields A and B in $\mathscr{D}$, denoted as $d_{\mathrm{A}\mathrm{B}}$, we first identified and counted the number of publications ($n_{\mathrm{A}\cup\mathrm{B}}$) with a positive field-affinity score for either or both fields A and B from the total of $N=100{,}700$ publications.
We then counted the number of publications ($n_{\mathrm{A}\cap\mathrm{B}}$) that had positive field-affinity scores for both fields A and B.
Finally, we calculated the ratio of these counts and subtracted it from 1 to obtain the Jaccard distance between fields A and B:
\begin{equation}\label{distance:A}
d_{\mathrm{A}\mathrm{B}}(y)=1-\frac{n_{\mathrm{A}\cap\mathrm{B}}(y)}{n_{\mathrm{A}\cup\mathrm{B}}(y)}\,.
\end{equation}
A value of 0 indicates that fields A and B are identical, while a value of 1 signifies that they are maximally distant.

Once the $19\times 19$ distance matrix for $\mathscr{D}$ is obtained, we can calculate the interdisciplinarity index for each publication.
Specifically, labelling each publication by $i=1,\,\dots,\,N$ and denoting the set of publications from year $y$ as $S(y)$, the interdisciplinarity index for publication $i\in S(y)$ is calculated as follows:
\begin{equation}\label{div:i}
\mathit{\Delta}_{i}=\left(1-\sum_{\mathrm{A}\in\mathscr{D}}\sum_{\mathrm{B}\in\mathscr{D}}p_{i}^{\mathrm{A}}p_{i}^{\mathrm{B}}d_{\mathrm{A}\mathrm{B}}(y)\right)^{-1}\,.
\end{equation}
Here, we quantified interdisciplinarity using the so-called \emph{effective number} of disciplines \citep{Jost06,Leinster12,Okamura20}.
This index offers advantages over other commonly used diversity measures, such as the simple `variety' index, Simpson's index, or popular entropy measures like Shannon's, in that it not only accounts for the number of fields and their composition (i.e.~relative abundances) but also appropriately considers the distances between the constituent fields.
Moreover, it is the most universal, as it can mathematically accommodate both broad categorisation and finer subdivisions of constituents in a consistent manner, satisfying the so-called Nesting Principle \citep{Okamura20}, which is not upheld by other measures such as the Rao--Stirling index \citep{Rao82}.

Using the interdisciplinarity index for each publication $i$, as defined by Eq.~(\ref{div:i}), the interdisciplinarity of field $\mathrm{A}\in\mathscr{D}$ is calculated as follows.
First, for each publication year, the set of publications with a positive field-affinity score for field A is identified. Then, the mean of the $\mathit{\Delta}_{i}$ distribution, which is approximately bell-shaped, is calculated for field A within this set, and this value is taken as the interdisciplinarity index for field A:
\begin{equation}\label{div:A}
\mathit{\Delta}_{\mathrm{A}}(y)=\sum_{i\in S(y)}\mathit{\Delta}_{i}P(\mathit{\Delta}_{i})\,,
\end{equation}
where $P(\mathit{\Delta}_{i})$ denotes the probability of the interdisciplinarity variable taking the value $\mathit{\Delta}_{i}$ in the distribution.
Using this index, we trace the evolution of each field's interdisciplinarity, i.e., the degree to which it is associated with other fields.

\subsection{The SDG-affinity score}

In addition to the field-affinity scores, our compiled OpenAlex data also contained a score indicating the affinity of each publication to each of the 17 SDGs, which we refer to as the SDG-affinity score.\footnote{\url{https://docs.openalex.org/api-entities/works/work-object\#sustainable_development_goals}.}  
The scores are calculated using specific machine learning algorithms \citep{Jaworeck22,Vanderfeesten21}.  
This score is assigned to each publication for each SDG, labelled by $m=1,\,\dots,\,17\,(\equiv M)$, as a number between 0 and 1.  
A score of 0 indicates that a publication is not related to the SDG, while a score of 1 indicates maximal relevance to the SDG.

To investigate the time evolution of the relative contribution of each field to each SDG, we first identify a set of publications, labelled by $i$, with positive affinity scores ($a_{i}>0$) for each SDG $m$ in each publication year ($y$).  
We then calculate the weighted sum of the affinity scores for each field ($\mathrm{A}\in\mathscr{D}$) across all publications belonging to this set, using the number of citations $(c_{i})$ for each publication as the weight, and denote this as $a_{\mathrm{A},m}(y)$.
Specifically, denoting the set of publications with a positive `field A'-affinity score as $S(\mathrm{A})$ and those with a positive `SDG $m$'-affinity score as $S(m)$, we have:
\begin{equation}
a_{\mathrm{A},m}(y)=\sum_{i\in S(y)\cap S(\mathrm{A})\cap S(m)}c_{i}a_{i}\,.
\end{equation}
In the above, the rationale for using the number of citations as a weight is based on the assumption that the attentional impact of a publication is proportionally associated with how many times it is subsequently cited.
Then, the share of contribution from field $\mathrm{A}$ to SDG $m$ in year $y$ is evaluated as:
\begin{equation}\label{ratio:A,m}
R_{\mathrm{A},m}(y)=\frac{a_{\mathrm{A},m}(y)}{\sum_{m=1}^{M}a_{\mathrm{A},m}(y)}\,,
\end{equation}
satisfying $\sum_{m=1}^{M}R_{\mathrm{A},m}(y)=1$ for each year $y$.
This index provides indirect insights into the relative increase (or decrease) in contributions from various fields, by publication year and by SDG.

\section{Results and discussion: The shifting nexus of SDGs and disciplines\label{sec:result}}

\subsection{Disciplinary boundaries: Convergence with or isolation from other fields}

\begin{figure*}[!t]
\centering
\includegraphics[width=0.94\textwidth]{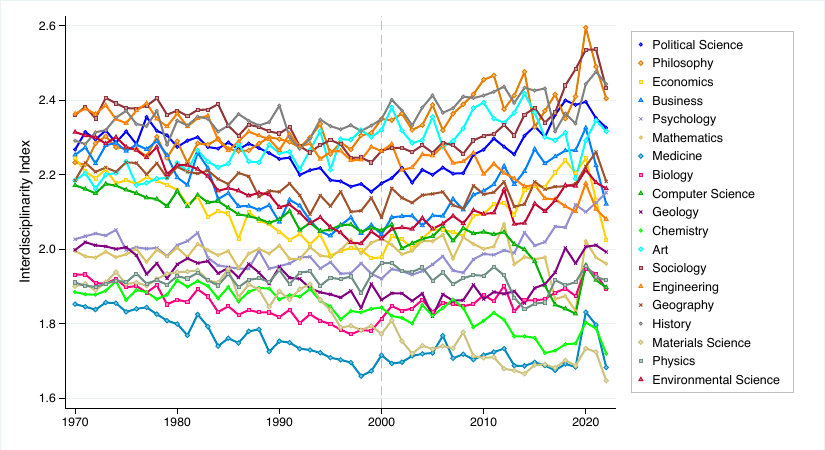}
\caption{\textbf{Trends in interdisciplinarity across fields.}
For each field, a higher interdisciplinarity index score, calculated using Eq.~(\ref{div:A}), indicates a greater degree of interrelation with other fields. [Source: OpenAlex]}\label{fig:2}
\end{figure*}

Before examining the contributions of different fields to specific global societal challenges, we first present an analysis of how the degree of disciplinary convergence for each field, as measured by the index defined in Eq.~(\ref{div:A}), has evolved over time.
The results of this analysis are shown in Figure \ref{fig:2}.
From the graph, we observe significant differences in the level of disciplinary convergence, as indicated by the interdisciplinarity index, across various fields.
However, the most striking finding is not just the variance between fields, but the common trend seen across many: around the year 2000 (indicated by the dotted line), there is a distinct shift in the trajectory of the interdisciplinarity index for many fields.

From around 1970 to 2000, although variations existed among fields, many showed a declining trend in interdisciplinarity.
In fact, according to a simple linear regression analysis, the interdisciplinarity indices declined in 14 out of 19 fields ($p<0.001$).
This may reflect a growing tendency for highly cited papers to become increasingly specialised within their respective disciplines, with mechanisms for generating new knowledge becoming more confined to these disciplines.
In contrast, after around 2000, the declining trend in interdisciplinarity appeared to cease in many fields, and an overall increase in interdisciplinarity was observed across these fields.
In fact, interdisciplinary indices rose in 10 out of 19 fields ($p<0.001$).
The early 2000s coincide with the formulation of the MDGs at the UN and the initiation of related initiatives.
The goals set at that time were difficult to achieve through the efforts of a single discipline alone and required collaboration across disciplines.

Assessing how directly this shift in interdisciplinarity indices reflects trends related to the global momentum associated with the MDGs and SDGs is complex.
However, beyond these UN initiatives, Figure \ref{fig:2} strongly suggests that the global community, including academia, has increasingly introduced interdisciplinary perspectives, interests, or approaches into discussions.
This phenomenon may also reflect the active publication of interdisciplinary research by high-impact journals and the proliferation of policy papers and other publications dealing with interdisciplinary issues.
These observations are also consistent with the sharp increase in `interdisciplinarity'-related publications in Figure \ref{fig:1} since the 2000s.

\subsection{Evolution of interdisciplinary contributions across SDGs}

\begin{figure*}[!p]
\centering
\includegraphics[width=0.93\textwidth]{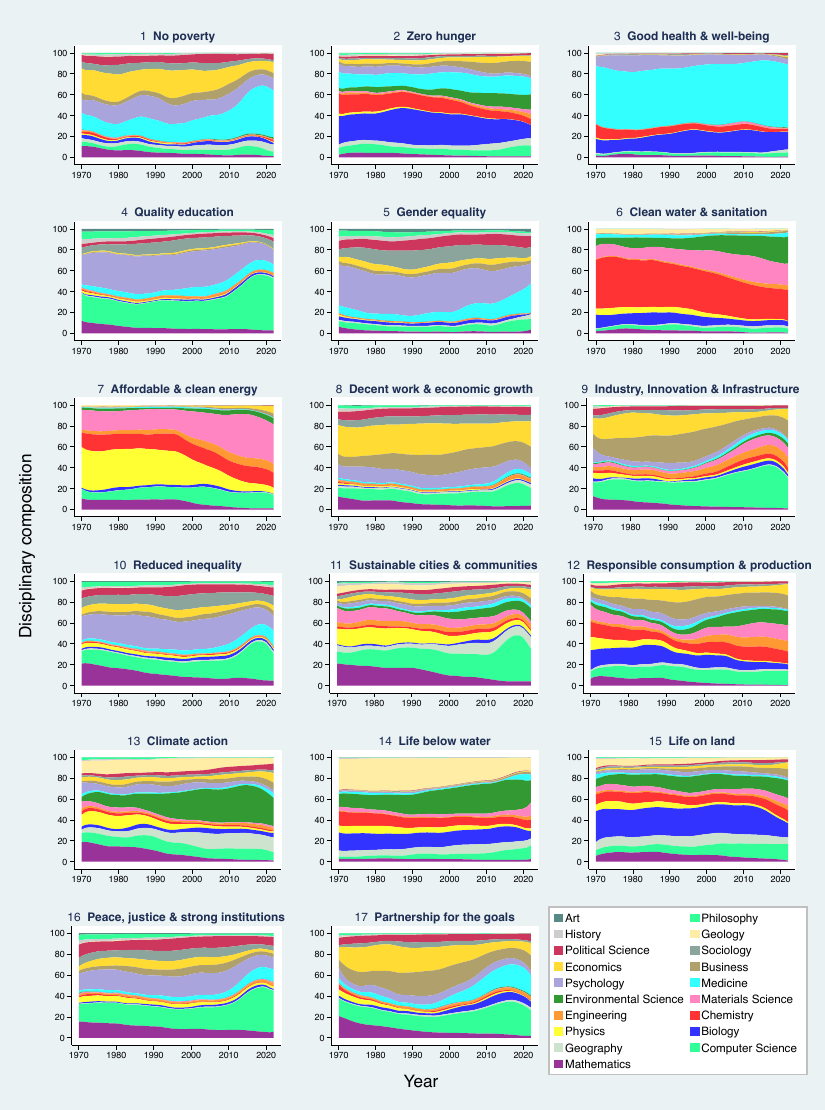}
\caption{\textbf{Evolution of interdisciplinary contributions across the SDGs.}
For each SDG, the share of each field is calculated using Eq.~(\ref{ratio:A,m}). [Source: OpenAlex]}\label{fig:3}
\end{figure*}

Building on the observed trend of disciplinary convergence across various fields, we now examine the interdisciplinary efforts directed towards each SDG and how the disciplinary composition has evolved over time through the analysis of the index defined in Eq.~(\ref{ratio:A,m}).
The results are illustrated in Figure \ref{fig:3}.
This figure demonstrates that different global challenges are closely linked to knowledge from a wide range of fields.
Furthermore, the degree of this linkage is unique to each challenge and has shifted dynamically over time.
In some instances, the diversity of contributing fields has increased, while in others, the dominance of certain fields has grown, varying across different periods and disciplines.

For instance, in the case of SDG 3 (\textsc{Good Health and Well-Being}), medicine has consistently accounted for over 50\% of the contributions throughout the entire period, while biology has maintained approximately 20\%.
Although there is some degree of interdisciplinarity, this goal is predominantly influenced by these few key fields.
This observation aligns with findings from an earlier study \citep{Makarenko21}, which noted that interest in SDG 3 within academic circles is primarily focused on medical aspects, while economic dimensions are poorly represented.
According to the cited literature, this relates to the underlying reasons for the lack of support and sufficient empirical evidence needed to address the financing challenges hindering the achievement of SDG 3.

In contrast, SDG 4 (\textsc{Quality Education}) has witnessed a significant increase in contributions from computer science over the past decade, despite its relatively steady presence in previous years.
This trend may reflect ongoing debates regarding the impact of information and communication technology (ICT) and emerging digital technologies, such as Big Data, AI, and the Internet of Things (IoT), on sustainable education \citep{Ghanem20,Tyagi19}.
This surge in prominence for computer science is also observed in other SDGs, including SDG 9 (\textsc{Industry, Innovation, and Infrastructure}), SDG 10 (\textsc{Reduced Inequalities}), SDG 11 (\textsc{Sustainable Cities and Communities}), SDG 16 (\textsc{Peace, Justice and Strong Institutions}) and SDG 17 (\textsc{Partnerships for the Goals}).
Given the rapidly evolving discussions surrounding the active use of AI, particularly in relation to achieving the SDGs \citep{Gupta21,Nastasa24,Singh24,UN24}, it is likely that the contributions from information science to many SDGs---or their revised goals, if any, to be defined as successors---will continue to grow in the future, regardless of whether this is evident in graphs such as Figure \ref{fig:3}.

For SDG 5 (\textsc{Gender Equality}), psychology accounted for approximately 40\% of contributions until the mid-2000s, but this figure has since declined to around 20\%.
In contrast, the share of contributions from medicine has increased.
While it is challenging to pinpoint the specific drivers behind this trend, discussions regarding the need to develop and strengthen health systems to support the women's empowerment aspect of SDG 5 may have contributed.
Furthermore, it is possible that medicine has partially merged with psychology to create a new interdisciplinary field relevant to SDG 5, which warrants further examination through expert knowledge.
The consistent 10\% contribution from political science throughout the period also reflects the political factors and interests surrounding this goal.

For SDG 7 (\textsc{Affordable and Clean Energy}), physics accounted for 30--40\% of contributions before 2000, but its share has since declined to around 6\%.
In its place, materials science has gained prominence, accounting for approximately 40\% of contributions since 2010.
Chemistry has consistently contributed around 15\%, while computer science has maintained a 15\% share since 2000.
This shift in the disciplinary composition may indicate a transition from a phase of scientific discovery to one more focused on engineering and practical applications in addressing energy issues.
Other SDGs also offer distinct insights, each reflecting specific disciplinary characteristics as well as potential historical, social, cultural, industrial, or political contexts.

Comparing the panels in Figure \ref{fig:3}, one observes that some SDGs exhibit an overall increasing trend towards disciplinary diversity, while others do not.
To investigate this further, we conducted an additional analysis to examine the changes in the level of interdisciplinarity associated with each SDG.
Specifically, for each SDG and each publication year, we identified publications with an SDG-affinity score above a certain threshold---here, a score of 0.5 was selected---and calculated the weighted mean interdisciplinarity index score for this set, using the number of citations for each publication as the weighting factor.
The rationale for this approach remains consistent; in other words, the citation count serves as a proxy variable for attention.
It is also important to note that this index abstracts the specific combinations and compositions of fields.

\begin{figure}[!p]
\centering
\includegraphics[width=0.93\textwidth]{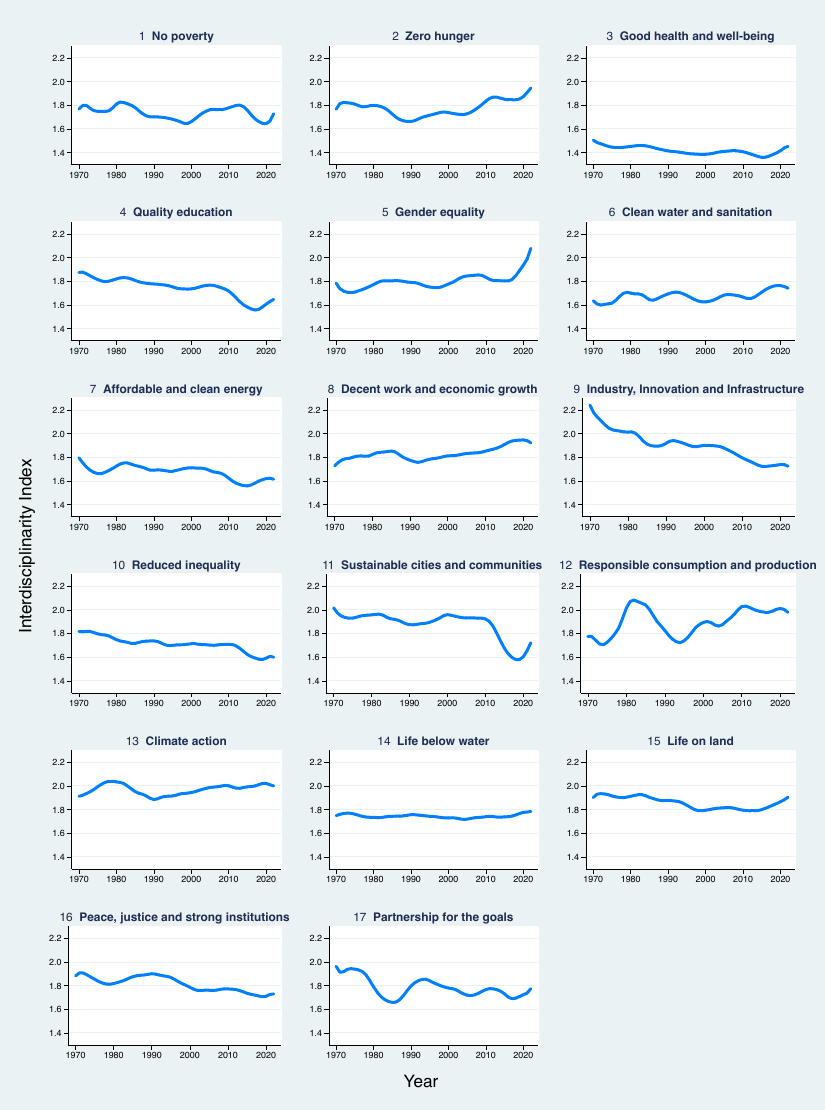}
\caption{\textbf{Trends in the Interdisciplinarity Index across the SDGs.}
For each SDG, the interdisciplinarity index is calculated based on its disciplinary composition, as depicted in Figure \ref{fig:3}, with appropriate weighting, as well as the distances between fields, which are calculated using Eq.~(\ref{distance:A}). [Source: OpenAlex]}\label{fig:4}
\end{figure}

The results are displayed across all SDGs in Figure \ref{fig:4}, revealing some intriguing trends characteristic of each SDG.
For instance, as mentioned earlier, SDG 3, predominantly influenced by medicine, results in a consistently low interdisciplinarity index.
In contrast, SDG 13 (\textsc{Climate Action}) has maintained a high level of disciplinary diversity, reflecting contributions from various fields, including environmental science, geography, geology, computer science, and physics.
Meanwhile, SDG 9 (\textsc{Industry, Innovation, and Infrastructure}), which was influenced by fields such as mathematics and psychology in the 1970s, has recently seen a sharp rise in the prominence of computer science, leading to a marked decline in its interdisciplinarity index.

\section{Conclusions: Towards evidence-based actions for sustainable development\label{sec:concl}}

This paper has provided an overview of how various disciplines are connected to different global societal challenges and how these connections have evolved over more than half a century, an area that has previously lacked substantial quantitative evidence.
Specifically, we employed two types of scores based on specific machine learning algorithms in the OpenAlex platform: field-affinity scores and SDG-affinity scores.
These measures allowed us to analyse the organic and dynamic relationships between different disciplines and the SDGs.

Our research question asked how the connections between research disciplines and global societal challenges---here represented by the SDGs---have developed over time.
Our analysis provides a clear answer: since the 2000s, coinciding with the implementation of the MDGs, interdisciplinary publications and citations have increased markedly, with many fields showing a tendency to become more interconnected.
At the same time, the evolution of interdisciplinarity has varied significantly across SDGs, reflecting the unique characteristics of each challenge.
In this sense, the study contributes novel quantitative evidence of both the overall strengthening of interdisciplinarity and the heterogeneous dynamics shaping specific global issues.

Our analysis also reveals striking asymmetries across specific SDGs.
For example, SDG3 has consistently been dominated by Medicine, whereas other goals reflect more balanced contributions from a wide range of disciplines.
Moreover, the degree of interdisciplinarity itself shows diverse trajectories, with some SDGs exhibit increasing integration over time, others show little change, and a few even display declining trends, reflecting the distinctive characteristics of each societal challenge.
These extreme results underscore the heterogeneous and dynamic nature of interdisciplinarity: while certain global issues are addressed primarily through the lens of a single discipline, others require, and indeed demonstrate, a more diversified set of disciplinary engagements, with compositions that continue to evolve over time.

Beyond its scholarly contribution, the study also offers important practical implications.
Tackling complex societal challenges requires not only the integration of the latest scientific knowledge and emerging technological trends, such as the rapid incorporation of AI, but also effective responses to major initiatives spearheaded by influential countries and international organisations.
These findings underscore the need for funding agencies and policymakers, working closely with international research communities, to sustain and expand efforts that foster interdisciplinarity.
Such efforts should be internationally coordinated and supported by continually updated technological infrastructures, ensuring that the growing interdisciplinarity observed in academic research translates into more effective real-world solutions.

In particular, funding agencies and policymakers may consider three complementary measures:
(\textit{i}) establishing dedicated funding programmes that explicitly target interdisciplinary research related to the SDGs---and their successors in the post-2030 agenda---in the era of advanced AI platforms;
(\textit{ii}) issuing calls for proposals that require or strongly encourage cooperation across fields, including the Arts, Humanities, and Social Sciences, where appropriate; and
(\textit{iii}) incentivising cross-border partnerships between academia, industry, and policy sectors through effective and future-oriented science diplomacy where needed.
Taken together, these measures would help ensure that the growing interdisciplinarity observed in academic research is effectively channelled into real-world solutions to global societal challenges.

As we approach the five-year mark before the SDG target deadline, the insights and visualisations presented here provide valuable perspectives from an interdisciplinary and cross-sectoral standpoint.
They offer both researchers and decision-makers a clearer understanding of how disciplinary boundaries have shifted in response to global challenges, while also highlighting concrete pathways for advancing collaboration and evidence-based action in the years to come.

Finally, this study highlights several methodological considerations and limitations that also point to directions for future research.
Because the analysis relies on OpenAlex, the findings inevitably reflect the characteristics of this database.
Although coverage is expanding, it remains uneven across disciplines, regions, and languages: English-language outputs are generally overrepresented, while publications in local languages or regional journals are less visible.
Nevertheless, compared with other major databases, OpenAlex offers relatively better representation of non-English publications and has advantages in several other respects \citep{Priem22,Delgado-Quiros24}.
In particular, it provides complementary strengths to commercial databases, especially for policy-related perspectives where quick and broad overviews are essential \citep{Kitajima25,OECDnd,Okamura23,Okamura24,UNESCO23}.
Recognising the unique features of each database, including their strengths and limitations, and combining them appropriately can generate richer insights into global research dynamics \citep{Culbert25,Haupka24,Liu23,Visser21}.

A further limitation concerns the field-classification scheme employed, since alternative specifications could yield different distributions of interdisciplinarity and collaboration.
Acknowledging these constraints clarifies the scope of the present contribution and highlights the importance of integrating complementary data sources and refining classification methods in future work.
In this way, bibliometric analyses can continue to evolve as a crucial tool for examining the interplay between research and global societal challenges, thereby strengthening both scholarly understanding and policy-oriented debates.
By fostering such methodological refinement, future work can further enhance the quality and dynamism of discussions on societal impact in both policy and academic spheres, contributing meaningfully to these critical dialogues.

\vspace{2mm}
\paragraph{\textbf{Acknowledgements.}}
The views and conclusions contained herein are those of the author and should not be interpreted as necessarily representing the official policies or endorsements, either expressed or implied, of any of the organisations with which the author is currently or has been affiliated in the past.


\vspace{-0.5em}
\paragraph{\textbf{Competing Interests.}}
The author has no competing interests.

\vspace{-0.5em}
\paragraph{\textbf{Funding Information.}}
The author did not receive any funding for this research.

\vspace{-0.5em}
\paragraph{\textbf{Data Availability.}}
The datasets and figures generated and/or analysed during this study are available in the Zenodo repository at \url{https://doi.org/10.5281/zenodo.13998979}.

\renewcommand*{\bibfont}{\small}
\setlength{\bibsep}{0.2\baselineskip plus 0.2\baselineskip}
\def\bibfont{\footnotesize}

\newpage
\appendix

\pagestyle{fancy}
\fancyhead[LE,RO]{\textcolor{orange}{\footnotesize{\textsf{SUPPLEMENTARY MATERIALS}}}}
\fancyhead[RE,LO]{}
\fancyfoot[RE,LO]{\color[rgb]{0.04, 0.73, 0.71}{}}
\fancyfoot[LE,RO]{\scriptsize{\textbf{\textsf{\thepage}}}}
\fancyfoot[C]{}

\renewcommand{\thefigure}{S\arabic{figure}}
\renewcommand{\thetable}{S\arabic{table}}
\renewcommand{\theequation}{S\arabic{equation}}
\renewcommand{\headrule}{\color{orange}\oldheadrule}

\setcounter{section}{0}
\setcounter{subsection}{0}
\setcounter{figure}{0}
\setcounter{table}{0}
\setcounter{equation}{0}

\section*{Supplementary Materials}

The data derived from OpenAlex and used in this study, as outlined in the Methods section of the main text, are available in the Zenodo repository at \url{https://doi.org/10.5281/zenodo.13998979}.
Figure \ref{fig:1} is based on the data presented in Table \ref{tab:1}, while Figures \ref{fig:2}, \ref{fig:3}, and \ref{fig:4} are derived from the data in Table \ref{tab:2}.

\addcontentsline{toc}{section}{Supplementary Materials}

\begin{table}[!h]
\vspace{2em}
\caption{\textbf{The data contained in the `{Suppl\_data\_1.csv}' file.}
Domains are labelled by $\texttt{\textcolor{red}{$\alpha$}}=\texttt{1},\,\dots,\,\texttt{4}$, where \texttt{1}: Life Sciences, \texttt{2}: Social Sciences, \texttt{3}: Physical Sciences, and \texttt{4}: Health Sciences.}
\label{tab:1}
\centering
{\small
\begin{tabular*}{\linewidth}{@{} p{5em} p{41em}}
\toprule
~Variable & Description \\
\midrule
~{\texttt{pyear}} & Year of publication \\
~{\texttt{nwork}} & Number of publications, indexed in OpenAlex \\
~{\texttt{nwork\textcolor{red}{$\alpha$}}} & Number of publications in domain \texttt{\textcolor{red}{$\alpha$}}, indexed in OpenAlex \\
~{\texttt{nIDR}} & Number of publications with `multi-/inter-/trans-disciplinar*' in Title or Abstract, indexed in OpenAlex \\
~{\texttt{nIDR\textcolor{red}{$\alpha$}}} & Number of publications in domain \texttt{\textcolor{red}{$\alpha$}} with `multi-/inter-/trans-disciplinar*' in Title or Abstract, indexed in OpenAlex \\
~{\texttt{{\%}nIDR}} & Ratio of \texttt{nIDR} to \texttt{nwork} (in \%) \\
~{\texttt{{\%}nIDR\textcolor{red}{$\alpha$}}} & Ratio of \texttt{nIDR\textcolor{red}{$\alpha$}} to \texttt{nwork\textcolor{red}{$\alpha$}} (in \%) \\
\bottomrule
\end{tabular*}
}
\end{table}
\vspace{1em}
\begin{table}[!h]
\caption{\textbf{The data contained in the `{Suppl\_data\_2.csv}' file.}
Disciplines are labelled by $\text{\textcolor{red}{k}}=1,\,\dots,\,19$, where
\texttt{1}: Political Science,	
\texttt{2}: Philosophy,			
\texttt{3}: Economics,			
\texttt{4}: Business,			
\texttt{5}: Psychology,			
\texttt{6}: Mathematics,		
\texttt{7}: Medicine,			
\texttt{8}: Biology,			
\texttt{9}: Computer Science,	
\texttt{10}: Geology,			
\texttt{11}: Chemistry,			
\texttt{12}: Art,				
\texttt{13}: Sociology,			
\texttt{14}: Engineering,		
\texttt{15}: Geography,			
\texttt{16}: History,			
\texttt{17}: Materials Science,	
\texttt{18}: Physics,			
and \texttt{19}: Environmental Science.	
}
\label{tab:2}
\centering
{\small
\begin{tabular*}{\linewidth}{@{} p{5em} p{41em}}
\toprule
~Variable & Description \\
\midrule
~{\texttt{idwork}} & Work identifier starting with `W', as defined in OpenAlex \\
~{\texttt{pyear}} & Year of publication \\
~{\texttt{citation}} & Number of citations, as of April 2024, indexed in OpenAlex \\
~{\texttt{discip\textcolor{red}{k}}} & Score of affinity to each of the level-0 concepts defined in OpenAlex ($\text{\textcolor{red}{k}}=1,\,\dots,\,19$) \\
~{\texttt{SDG\textcolor{red}{m}}} & Score of affinity to each of the United Nations' Sustainable Development Goals (SDGs) ($\texttt{\textcolor{red}{m}}=\texttt{1},\,\dots,\,\texttt{17}$) \\
\bottomrule
\end{tabular*}
}
\end{table}

\end{document}